# Design study of magnetic environments for XYZ polarization analysis using ³He for the new thermal time of flight spectrometer TOPAS

Zahir SALHI, Earl BABCOCK, Alexander IOFFE
Jülich Centre for Neutron Science at the FRM II, Garching, Germany

**Abstract:**
We present a finite element calculation of the magnetic field (MagNet software) taken with the newly proposed PASTIS Coil, which uses a wide-angle banana shaped 3He Neuton Spin Filter cell (NSF) to cover a large range of scattering angle. The goal of this insert is to enable XYZ polarization analysis to be installed on the future thermal time-of flight spectrometer TOPAS.

**Introduction:**
Polarization analysis, PA, of polarized neutrons is a powerful tool for separation of nuclear spin-incoherent background, analysis of complex magnetic structures and the study of magnetic excitations. Several wide angle spectrometers with polarization analysis exist or are under construction in which PA is used. The PA can be performed in a variety of ways depending on the instrument's parameters, but with performance limited by the analyzer height and integration over the height of the detectors. Installation of a new longer, height-position sensitive detector bank gives a unique opportunity to prototype and test a polarized ³He XYZ analysis system which could utilize the full height and position resolution of these new detectors. We present an initial design study with finite element magnetic field (FEM) calculations of possible XYZ field configurations suitable for polarized 3He and adapted to the DNS instrument geometry. Two clear options exist, a magnetized mu-metal geometry, similar to ref [1], or a resistive coil set similar to ref [2], however in our proposed designs, certain key differences exist which build on the experiences from prior devices.

**Helium-3 relaxation time:**
The Helium-3 polarization in NSF cell decays to thermal equilibrium with characteristic time constant $T_1$. The longitudinal relaxation rate $\Gamma_1 = T_1^{-1}$ consists on three main mechanisms for relaxations of spins and characterized by:

$$\Gamma_1 := \frac{1}{T_1} = \frac{1}{T_{1\,dd}} + \frac{1}{T_{1_G}} + \frac{1}{T_{1\,wall}} \quad (1)$$

$T_{1\,dd}$ is the dipole-dipole relaxation time due to interaction between colliding ³He atoms. The magnetic dipole coupling of atoms during a binary collision results in the loss of nuclear polarisation. This rate is well known and described by:

$$\frac{1}{T_1^{dipole}} = \frac{p}{\text{bar}} \cdot \frac{1}{817\text{h}} \quad (2)$$

Where P is given the pressure an h means hours, this relaxation effect is only relevant at high pressure limiting long storages (for P=2,7 bar the $T_1^{dipole}$ = 300h)

$T_{1_G}$ is the relaxation time due to the diffusion of ³He through magnetic field gradient:

$$T_{Grad}[\text{h}] = \frac{7000}{P[\text{bar}]} \left( \left(\frac{\Delta B_X}{B}\right)^2 + \left(\frac{\Delta B_y}{B}\right)^2 \right) \quad (3)$$

In the presence of magnetic field gradient, a moving atom will experience different magnetic field strengths. If the field fluctuations are close to the transition frequency between Zeeman energy level, then spin flip can be inducted causing a loss of magnetization.

$$\left(\frac{\partial B_r / \partial r}{B_0}\right) \leq 10^{-3} \tfrac{1}{cm} \rightarrow T_1^{grad} \geq 100h \rightarrow P = 2,7 bar$$

The final term, $T_{1\ wall}$ is the relaxation time due to the interactions with the cell wall and is the least understood $^3$He relaxation mechanism, this relaxation depends purely on the characteristics of the container which can contains magnetic impurities like magnetite (most common impurities in the glass), absorption of helium into the glass may also cause relaxation by precession of the helium moment about the net local gradient field caused by the magnetic impurities. This phenomena is observed when the cells are used in high magnetic field.

**Design:**
A large step has been made in the design of wide angle polarization analysis. Recently the ILL has developed a wide angle polarization analysis prototype called PASTIS, where a set of x, y and z Helmholtz coils was used to create a homogeneous magnetic field in an arbitrary direction [1]. The PASTIS design has been further developed (Fig.1) introducing mu-metal sheets to produce an improved homogeneous magnetic field in a large region and to reduce the blind area due to the support material of the Helmholtz coils. The center of the configuration is the sample, placing the $^3$He cell significantly off center. The present design has nearly a 45° window in the vertical direction, providing polarization analysis for the full height of the detector.

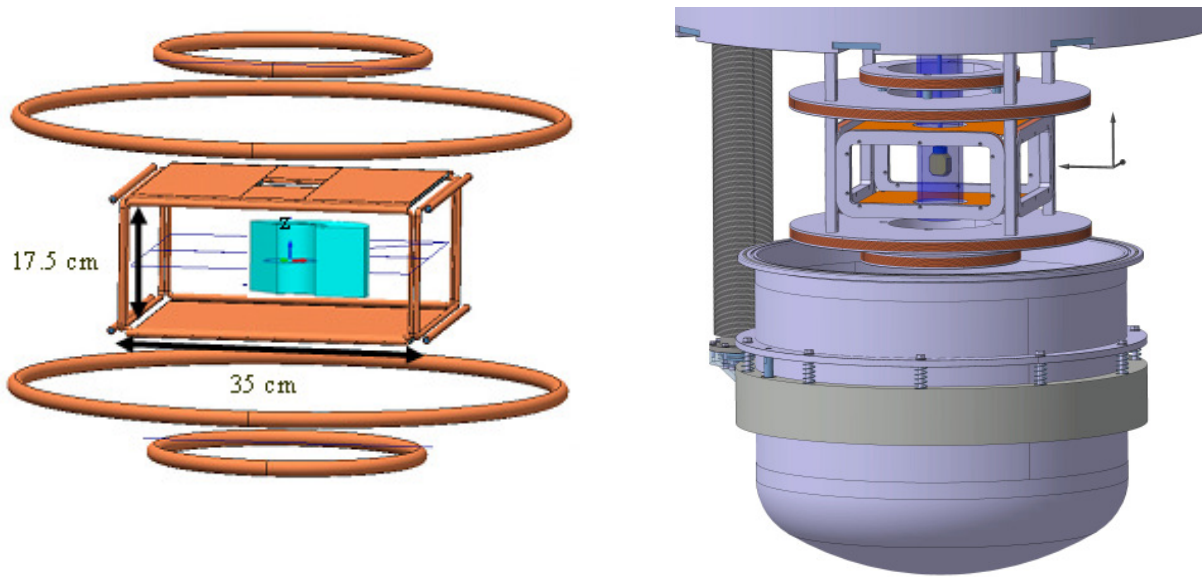

Fig. 1: Coil layout for the PASTIS concept.

To keep the depolarization of the polarized 3He sufficiently low it is necessary that the field gradient is below $10^{-4}$ (equation 3) over the volume of a large NSF Cell. Thereby allowing the cells to remain on the instrument longer and eliminating refilling and transport issues in short time. Fig. 2 show finite element calculations of the created magnetic field in the x,y and z direction . It is clear that the inner part is highly homogeneous, including also the position of the $^3$He cell. As seen from Fig. 1 the coil setup fulfills this condition for a large area, allowing the polarization analysis in a wide angular range. We are constructing a prototype of this new setup. In parallel the construction of the polarizer based on the design of the MARIA reflectometer in continuing. Once TOPAS is going to operation it will provide polarization analysis covering a wide angular range from day 1.

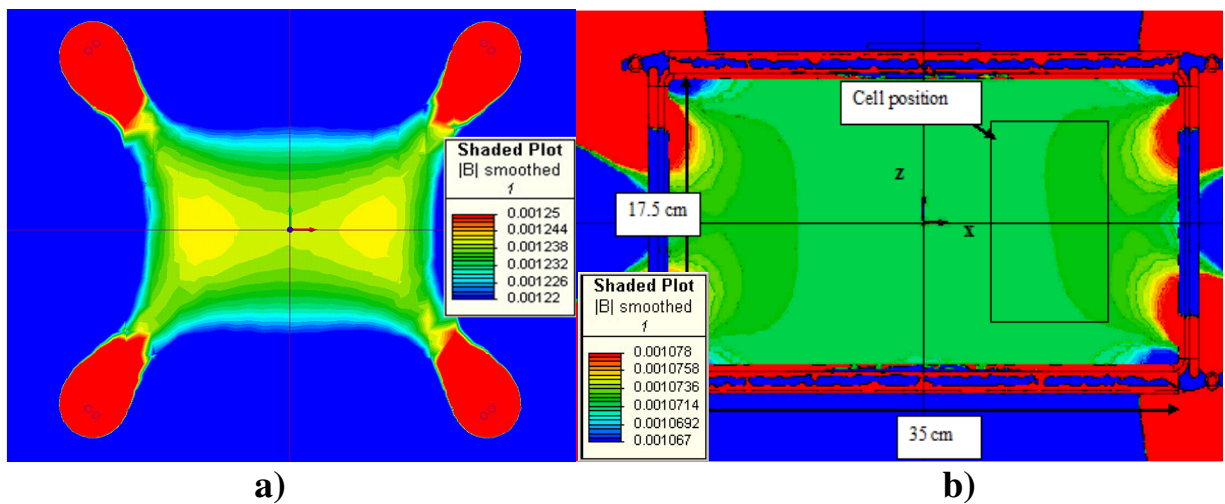

Fig. 2: a) A contour plot of the calculated Magnetic field distribution in the x, y direction.
b) A contour plot of the calculated Magnetic field distribution in the z direction highlighting the homogeneity of the magnetic field

E-mail of the corresponding author: e.babcock@fz-juelich.de, z.sahli@fz-juelich.de